\documentclass[amsmath,prb,twocolumn,superscriptaddress]{revtex4}
\usepackage{color}
\usepackage{amsfonts}
\usepackage{amssymb}
\usepackage{graphicx}
\usepackage{pstricks}
\usepackage{braket}

\begin{document}
\title{Nonequilibrium dual-boson approach}
\author{Feng Chen}
\email{fec011@ucsd.edu}
\affiliation{Department of Physics, University of California San Diego, La Jolla, CA 92093, USA}
\author{Mikhail I. Katsnelson}
\email{M.Katsnelson@science.ru.nl}
\affiliation{Radboud University Nijmegen, Institute for Molecules and Materials, 6525AJ Nijmegen, The Netherlands}
\author{Michael Galperin}
\email{migalperin@ucsd.edu}
\affiliation{Department of Chemistry \& Biochemistry, University of California San Diego, La Jolla, CA 92093, USA} 

%%%%%%%%%%%%%%%%%%%%%%%%%%%%%%%%%%%%%%%%%%%%%%%%%%%%%%%%%%%%%%%%%%%%%%%%%%%%%%%
\begin{abstract}
We develop nonequilibrium auxiliary quantum master equation dual boson method (aux-DB),
and argue that it presents a convenient way to describe steady states of correlated impurity models
(such as single molecule optoelectronic devices)
where electron and energy transport should be taken into account. 
The aux-DB is shown to provide high accuracy with relatively low numerical cost.
Theoretical analysis is followed by illustrative simulations within
generic junction models, where the new scheme is benchmarked against
numerically exact results. 
\end{abstract}

\maketitle
%%%%%%%%%%%%%%%%%%%%%%%%%%%%%%%%%%%%%%%%%%%%%%%%%%%%%%%%%%%%%%%%%%%%%%%%%%%%%%%

\section{Introduction}\label{intro}
Fast development of nano-fabrication techniques combined with advances
in laser technology lead to tremendous progress in optical studies of
nanoscale systems. Optical spectroscopy of single molecules in
current carrying junctions became reality. 
Surface~\cite{ioffe_detection_2008,ward_simultaneous_2008,ward_vibrational_2011}
and tip~\cite{liu_revealing_2011,chiang_molecular-resolution_2015,lee_tip-enhanced_2017} 
enhanced Raman spectroscopies (SERS and TERS) 
as well as bias-induced 
electroluminescence~\cite{schneider_light_2012,cavar_fluorescence_2005,chen_viewing_2010,dong_generation_2010,imada_real-space_2016,imada_single-molecule_2017,kimura_selective_2019}
measurements
yield information on extent of heating of vibrational and electronic
degrees of freedom in biased junctions, electron transport noise 
characteristics, molecular structure, dynamics and chemistry.
Combination of molecular electronics with nonlinear optical spectroscopy
resulted in emergence of a new field of research coined 
optoelectronics~\cite{galperin_molecular_2012,galperin_photonics_2017}.

Optical response of single molecule junctions is only possible
due to strong enhancement of the signal by surface plasmons~\cite{gersten_electromagnetic_1980}.
Large fields and confinement result in strong interaction
between molecular and plasmonic excitations.
Note also recent experiments on ultra-strong light-matter
interaction in single molecule nano-cavities 
(at the moment, in the absence of electron 
current)~\cite{chikkaraddy_single-molecule_2016,kongsuwan_suppressed_2018}.
At nanoscale classical electrodynamics becomes inadequate
as it cannot describe quantum coherence and mixing
between plasmon and molecular exciton,
while strong interactions require to go beyond perturbation theory.

Development of theoretical methods for simulation of strongly 
correlated open nonequilibrium impurity systems is a preprequisite 
in modeling nanoscale molecular devices with potential applications
from optical characterization and control to energy harvesting, spintronics, and quantum computation.
With numerically exact techniques,
such as continuous time quantum Monte Carlo~\cite{cohen_taming_2015,antipov_currents_2017,ridley_numerically_2018}
or renormalization group methods~\cite{anders_steady-state_2008,schmitt_comparison_2010,schollwock_density-matrix_2005,schollwock_density-matrix_2011},  
being computationally costly and thus mostly focused on simple models,
relatively numerically inexpensive and
sufficiently accurate schemes for realistic simulations
are in high demand.

One of such perspective universal impurity solvers is 
the nonequilibrium dual fermion (DF)
approach originally introduced in 
Ref.~\onlinecite{jung_dual-fermion_2012}.
Recently, the approach was modified~\cite{chen_auxiliary_2019} 
to reduce computational cost and improve ability  
to simulate steady-states of correlated impurity models.
Note that focus of the dual fermion approach is electron transport.
At the same time, simulations of optoelectronic devices require accounting also
for energy transfer. 

Here, we introduce {\em auxiliary quantum master equation (QME) -
nonequilibrium dual boson (aux-DB) method} -
a universal nonequilibrium impurity solver
which accounts for both charge and energy transport in strongly
correlated open systems.
Similar to DF of Ref.~\onlinecite{jung_dual-fermion_2012} being
nonequilibrium version of the equilibrium DF method~\cite{rubtsov_dual_2008,hafermann_superperturbation_2009,antipov_opendf_2015,rohringer_diagrammatic_2018}
(DF inspired superperturbation theory),
aux-DB has its origin in equilibrium DB approach~\cite{rubtsov_dual_2012,van_loon_beyond_2014,van_loon_plasmons_2014,van_loon_ultralong-range_2015,stepanov_self-consistent_2016,van_loon_double_2016,stepanov_local_2016,stepanov_effective_2018}.
Below, after introducing nonequilibrium DB in Section~\ref{db}, in Section~\ref{aux} we present auxiliary 
quantum master equation (QME) treatment within the method.
Theoretical considerations are followed by illustrative numerical simulations
within generic junction models in Section~\ref{numres}.
Section~\ref{conclude} concludes. 

%%%%%%%%%%%%%%%%%%%%%%%%%%%%%%%%%%%%%%%%%%%%
\begin{figure}[b]
\centering\includegraphics[width=\linewidth]{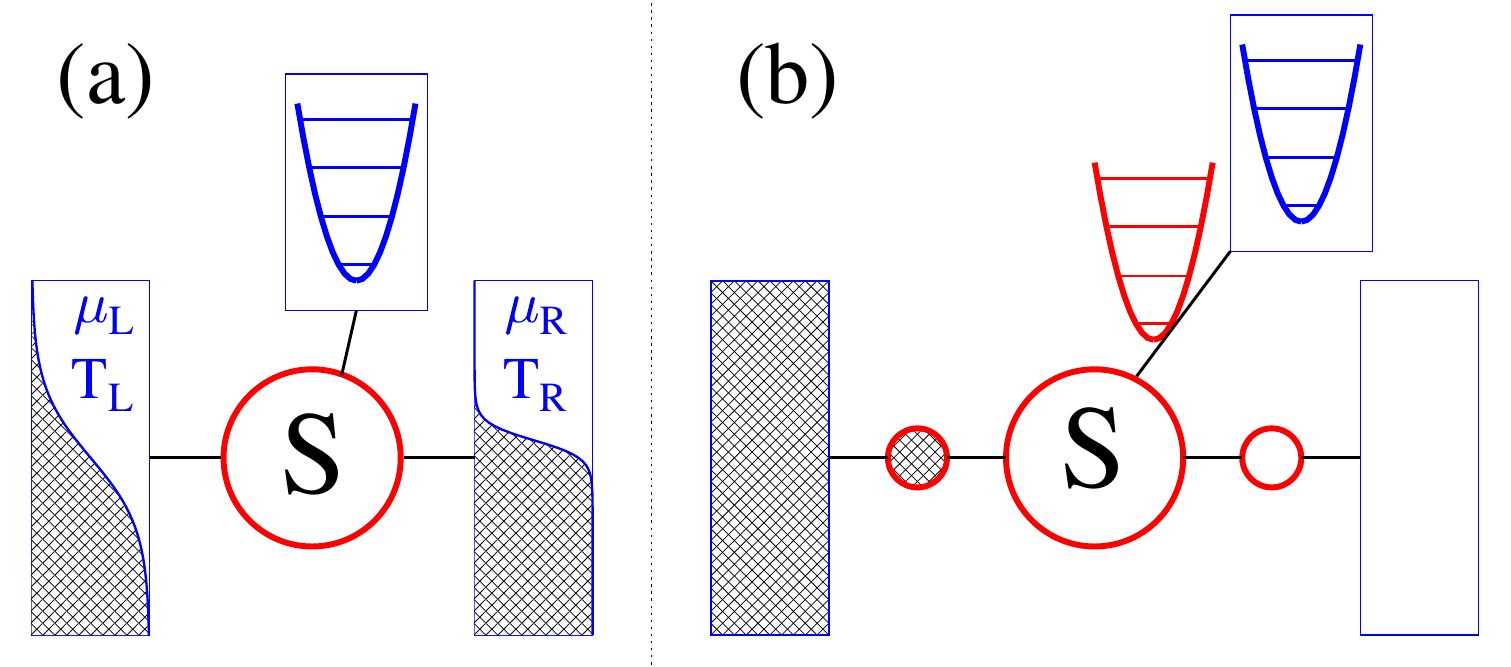}
\caption{\label{fig1}
Nonequilibrium junction model. Shown are
(a) Physical model and
(b) Reference system within aux-DB approach. 
}
\end{figure}
%%%%%%%%%%%%%%%%%%%%%%%%%%%%%%%%%%%%%%%%%%%%

%%%%%%%%%%%%%%%%%%%%%%%%%%%%%%%%%%%%%%%%%%%%%%%%%%%%%%
%%%%%%%%%%%%%%%%%%%%%%%%%%%%%%%%%%%%%%%%%%%%%%%%%%%%%%

% dual boson theory
\section{Nonequilibrium DB theory}\label{db}
Here we present a short description of the aux-DB method. Detailed derivations are given in 
Appendix~\ref{appA}.
Similar to the DF method, in the nonequilibrium DB  approach one considers reduced dynamics of an open quantum system
with interactions confined to the molecular subspace. Contrary to the DF method, in addition to contacts (Fermi baths)
the system is coupled also to Bose bath(s). Effect of the baths enters the effective action defined on the Keldysh
contour~\cite{kamenev_field_2011} via corresponding self-energies $\Sigma$ (for Fermi baths) and $\Pi$ (for Bose baths)
\begin{equation}
\label{S}
S[\bar d,d] = \bar d_1\,\big[G_0^{-1}-\Sigma^B\big]_{12}\, d_2
 - \bar b_1\,\Pi^B_{1,2}\, b_2 + S^{int}[\bar d,d]
\end{equation}
Here and below summation of repeating indices is assumed. 
In (\ref{S}) $\bar d_i\equiv d_{m_i}(\tau_i)$  ($d_i\equiv d_{m_i}(\tau_i)$) is the Grassmann variable corresponding to
creation (annihilation) operator $\hat d_{m_i}^\dagger(\tau_i)$ ($\hat d_{m_i}(\tau_i)$) 
represents both molecular (spin-)orbital $m_i$ and contour variable $\tau_i$, of an electron in orbital $m_i$ 
in the Heisenberg picture~\cite{negele_quantum_1988}.
$b_i = b_{m_1^{i}m_2^{i}}(\tau_i) \equiv \bar d_{m_1^i}(\tau_i) d_{m_2^i}(\tau_i)$ is the molecular excitation
representing optical transition within the molecule from orbital $m_2^i$ to orbital $m_1^i$ at contour variable $\tau_i$.
Sum over indices includes summation over molecular orbitals (optical transitions) and contour integration:
$\sum_i\ldots \equiv \sum_{m_i}\int_c d\tau_i\ldots$ ($\sum_{m_1^i,m_2^i}\int_c d\tau_i\ldots$).
$G_0^{-1}$ is the inverse free Green's function (GF)~\cite{wagner_expansions_1991}
\begin{equation}
\label{invG0}
\big[G_0^{-1}\big]_{12}\equiv \delta(\tau_1,\tau_2)\big[i\partial_{\tau_1}\delta_{m_1,m_2}
- H^0_{m_1m_2}(\tau_1)\big] -
\Sigma^{irr}_{12},
\end{equation}
$\Sigma^B$ and $\Pi^B$ are respectively self-energies due to coupling to Fermi (contacts) and Bose (plasmon) baths,
\begin{equation}
 \label{SP}
 \begin{split}
 &\Sigma^B_{m_1m_2}(\tau_1,\tau_2) =  V_{m_1k} g_k(\tau_1,\tau_2) V_{km_2}\quad \mbox{and}
\\
& \Pi^B_{m_1m_2,m_3m_4}(\tau_1,\tau_2) =  V_{m_1m_2,\alpha} d_\alpha(\tau_1,\tau_2) V_{\alpha,m_3m_4}.
\end{split}
 \end{equation} 
In Eqs.~(\ref{invG0})-(\ref{SP}),  $H^0_{m_1m_2}(\tau)$ is the non-interacting part of 
the molecular Hamiltonian, $\Sigma^{irr}_{m_1m_2}(\tau_1,\tau_2)\sim\delta(\tau_1,\tau_2)$ is 
the irregular self-energy, $V_{mk}$ and $V_{m_1m_2,\alpha}$ are matrix elements for electron transfer from contact state $k$ to
molecular orbital $m$ and  for optical electron transfer from orbital $m_1$
to $m_2$ with absorption of phonon in mode $\alpha$,  respectively.
$g_k(\tau_1,\tau_2)\equiv -i \langle T_c\,\hat c_k(\tau_1)\,\hat c_k^\dagger(\tau_2)\rangle$
and $d_\alpha(\tau_1,\tau_2)\equiv -i \langle T_c\,\hat a_\alpha(\tau_1)\,\hat a_\alpha^\dagger(\tau_2)\rangle$
 are GFs of free electron in state $k$ of the contacts
 and free phonon in mode $\alpha$.
 All intra-molecular interactions are within the (unspecified) contribution to the action $S^{int}[\bar d,d]$.

As in equilibrium DB~\cite{van_loon_beyond_2014}, one introduces an exactly solvable {\em reference system} (see below).
Similarly to aux-DF~\cite{chen_auxiliary_2019}, 
the true baths are approximated by a finite number of auxiliary discrete modes subject 
to Lindbladian evolution (see Fig.~\ref{fig1}b).
Thus, action of the reference system $\tilde S[\bar d,d]$ is known and has the same general form (\ref{S}) 
with true self-energies $\Sigma^B$ and $\Pi^B$ substituted by their approximate representations $\tilde \Sigma^B$ and $\tilde \Pi^B$.
The desired action $S$ can then be written as
\begin{equation}
\label{Srs}
S[\bar d,d]=\tilde S[\bar d,d]+ \bar d_1\,\delta\Sigma^B_{12}\, d_2
+ \bar b_1\,\delta\Pi^B_{12}\, b_2.
\end{equation}
where $\delta\Sigma^B\equiv\tilde\Sigma^B -\Sigma^B$ and $\delta\Pi^B\equiv\tilde\Pi^B-\Pi^B$.

Because direct application of standard diagrammatic expansion around the interacting reference system
is not possible (the Wick's theorem does not apply~\cite{fetter_quantum_1971}), 
two artificial particles, {\em dual fermion} ($f$) and {\em dual boson} ($\eta$), are introduced which is used to unravel 
last two terms in (\ref{Srs}) via the Hubbard-Stratonovich transformation~\cite{coleman_introduction_2015}.
Integrating out molecular fermions ($d$ and $\bar d$) and comparing 
the fourth order cumulant expansion of the resulting expression 
with the general form of action for dual particles,
\begin{equation}
\label{SD}
\begin{split}
S^{D}[f^{*},f] &= \bar f_1\, \big[\big(G_0^{DF})^{-1}-\Sigma^{DF}\big]_{12}\, f_2
\\
& + \bar \eta_1\, \big[\big(D_0^{DB})^{-1}-\Pi^{DB}\big]_{12}\, \eta_2,
\end{split}
\end{equation}
one gets  
\begin{equation}
\label{GDSP}
\begin{split}
&\big(G^{DF}_0\big)^{-1}_{12} = -g^{-1}_{12} - g^{-1}_{13}\,\big[\delta\Sigma^B\big]^{-1}_{34}\, g^{-1}_{42},
\\
&\big(D^{DB}_0\big)^{-1}_{12} = -\chi^{-1}_{12} - \chi^{-1}_{13}\,\big[\delta\Pi^B\big]^{-1}_{34}\, \chi^{-1}_{42},
\\
&\Sigma^{DF}_{12} = \bigg( \Gamma_{13;42} + 
i\big(\gamma_{514}\,\delta_{326}-\gamma_{512}\,\delta_{346}
\\&\qquad 
+\gamma_{532}\,\delta_{146}-\gamma_{534}\,\delta_{126}\big) \big[D^{DB}_0\big]_{65}\bigg) \big[G_0^{DF}\big]_{43}
\\ &\qquad
- \bigg( \langle \hat b_5^\dagger\rangle \chi^{-1}_{54}\gamma_{312} +\chi^{-1}_{35}\langle\hat b_5\rangle\delta_{124}\bigg) 
\big[D^{DB}_0\big]_{43}
\\
&\Pi^{DB}_{12} = -i\,\gamma_{145}\,\delta_{632}\, \big[G^{DF}_0\big]_{34}\, \big[G^{DF}_0\big]_{56}
\end{split}
\end{equation}
Here $g_{12}$ and $\chi_{12}$ are single particle GFs of fermion and molecular excitation 
of the reference system, 
$\gamma_{123}$, $\delta_{123}$ and $\Gamma_{13;24}$ are vertices~\cite{stefanucci_nonequilibrium_2013}
(see Eq.~(\ref{vertices}) and Fig.~\ref{figS1} in Appendix~\ref{appA}).

With dual particles GFs,
\begin{equation}
\begin{split}
\big(G^{DF}\big) &= \big[\big(G^{DF}_0\big)^{-1} -\Sigma^{DF}\big]^{-1}\quad\mbox{and}
\\ 
\big(D^{DB}\big) &= \big[\big(D^{DB}_0\big)^{-1} -\Pi^{DB}\big]^{-1},
\end{split}
\end{equation}
known, the single-particle ($G$) and two-particle ($D$) GFs of the molecule are obtained from
\begin{equation}
 \label{GD}
 \begin{split}
 G &=\big(\delta\Sigma^B\big)^{-1} + \big[g\,\delta\Sigma^B\big]^{-1}\, G^{DF} \big[\delta\Sigma^B\, g\big]^{-1}
 \\
 D &=\big(\delta\Pi^B\big)^{-1} + \big[\chi\,\delta\Pi^B\big]^{-1}\, D^{DF} \big[\delta\Pi^B\, \chi\big]^{-1}
 \end{split}
\end{equation}
Note, here the two-particle GF is correlation function of molecular optical excitation operators.
$G$ yields information on orbital populations, spectral functions and electron current in the junction,
while $D$ is used in calculation of boson (phonon) flux.

%%%%%%%%%%%%%%%%%%%%%%%%%%%%%%%%%%%%%%%%%%%%%%%%%%%%%%%

\section{Reference system}\label{aux}
Construction of a reference system to a large extent relies on accurate reproduction of the physical system's
hybridization functions $\Sigma^B$ and $\Pi^B$.
Accurate choice of the reference system parameters was recently discussed in
Refs.~\onlinecite{tamascelli_nonperturbative_2018,mascherpa_optimized_2019} for Bose baths 
and in Refs.~\onlinecite{arrigoni_nonequilibrium_2013,dorda_auxiliary_2015,dorda_optimized_2017,chen_markovian_2019} for Fermi baths.
Here we combine both considerations by introducing as the reference system 
physical system complemented with a  finite number of auxiliary unitary modes ($A$) subject to Lindbladian evolution.
This includes finite number of sites representing Fermi baths and modes
representing Bose bath (see Fig.~\ref{fig1}b  and Appendix~\ref{appB}). 
Dynamics of the extended $SA$ system (molecule plus finite number of sites and modes)
is driven by Markov Lindblad-type evolution
\begin{equation}
\label{qme}
 \frac{d\rho^{SA}(t)}{dt} = -i\mathcal{L}\rho^{SA}(t).
\end{equation}
Here, $\rho^{SA}(t)$ is the extended system density operator and $\mathcal{L}$ is the Liouvillian.
Note that Refs.~\onlinecite{tamascelli_nonperturbative_2018} and \onlinecite{chen_markovian_2019}
prove that, in principle, Markov dynamics of the extended system can exactly reproduce non-Markov unitary dynamics of the physical system $S$
as long as free correlation function of the auxiliary modes accurately reproduces the correlation function of 
the full baths.
However, in realistic calculations this representation is approximate due to restriction on number of auxiliary sites and modes
which can be taken in consideration. Thus, the aux-DB accounting for the difference between true and reference system
hybridization functions, Eq.~(\ref{GDSP}), is very useful in correcting the approximate mapping.
 
The aux-DB approach, Eqs.~(\ref{SD})-(\ref{GD}), requires single- and two-particle GFs 
$g$ and $\chi$ and vertices $\Gamma$, $\gamma$ and $\delta$ of the reference system as an input. 
Those are obtained by solving the QME~(\ref{qme}) and employing the quantum regression relation
(see Appendix~\ref{appC} for details).

Below we focus on steady state and consider a reference system of size small enough
that exact diagonalization can be employed.
For larger systems more advanced methods (e.g. matrix product 
states~\cite{dorda_auxiliary_2015}) may be used.
We note that while MPS is only works for 1d problems, this does not impose limitation
on the dimensionality of original (physical) problem, because any number and geometry
of couplings in the physical problem can be mapped onto effectively 1d formulation in auxiliary
reference system with only two (for Fermi) or one (for Bose) baths.

%%%%%%%%%%%%%%%%%%%%%%%%%%%%%%%%%%%%%%%%%%%%%%%%%%%%%%%

\section{Numerical results and discussion}\label{numres}
Here we illustrate the aux-DB method with numerical simulations within generic junction models:
resonant level model (RLM) and Anderson impurity model (AIM) coupled to Fermi and Bose baths.

% model
\subsection{Model}
We apply the aux-DB method to generic models with junction constructed from a system $S$ coupled to two
Fermi ($L$ and $R$) and one Boson bath ($P$) (see Fig.~\ref{fig1}a).
The Hamiltonian is
\begin{equation}
 \hat H = \hat H_S + \sum_{B=L,R,P}\big( \hat H_B+\hat V_{SB} \big),
\end{equation}
where 
\begin{equation}
\begin{split}
&\hat H_{L(R)} = \sum_{k\in L(R)}\varepsilon_k\, \hat c_{k}^\dagger\hat c_{k}
\\
&\hat H_{P} = \sum_{\alpha\in P}\omega_\alpha\hat a_\alpha^\dagger\hat a_\alpha
\end{split}
\end{equation}
are Hamiltonians of the contact $L$ ($R$) and phonon bath $P$.
\begin{equation}
\begin{split}
&\hat V_{SL\,(R)} =\sum_{m}\sum_{k\in L\,(R)}\big( V_{mk} \hat d_m^\dagger\hat c_{k} + H.c.\big)
\\
&\hat V_{SP} =\sum_{m_1,m_2}\sum_{\alpha\in P} V_{m_1m_2}^{\alpha} 
(\hat b_{m_1m_2}+\hat b_{m_1m_2}^\dagger) (\hat a_{\alpha} + \hat a_\alpha^\dagger)
\end{split}
\end{equation}
describe electron transfer  between the system and contact $L$ ($R$) and
 describes coupling to phonon $\alpha$ in the thermal bath $P$, respectively.
 Here, $\hat d_m^\dagger$ ($\hat d_m$) and $\hat c_{k}^\dagger$ ($\hat c_{k}$)
creates (annihilates) electron in orbital $m$ on the system and in state $k$ of the contacts, respectively,
$\hat a_\alpha^\dagger$ ($\hat a_\alpha$) creates annihilates phonon in mode $\alpha$, and
$\hat b_{m_1m_2}=\hat d_{m_1}^\dagger\hat d_{m_2}$.

For the system Hamiltonian we consider resonant level (RLM), 
\begin{equation}
\hat H_S=\varepsilon_0\,\hat n,
\end{equation}
and Anderson impurity (AIM), 
\begin{equation}
\hat H_S=\sum_{m=1,2}\varepsilon_0\,\hat n_m + U\hat n_1\hat n_2,
\end{equation}
models. Here, $\hat n_m=\hat d_m^\dagger\hat d_m$ and $U$ is the Coulomb repulsion.
In the AIM two types of coupling to the thermal bath are considered: 
symmetric, $V_{m_1m_2}^{\alpha}=\delta_{m_1,m_2}V_{m_1}^{\alpha}$, 
and anti-symmetric, $V_{m_1m_2}^{\alpha}=\delta_{m_1,m_2}(-1)^{m_1}V_{m_1}^{\alpha}$.

Using Eq.~(\ref{GD}) we calculate single- and two-particle GFs
and employ them to evaluate  the spectral functions $A_m(E)$, 
electron current~\cite{jauho_time-dependent_1994}, $I_L$, at the left interface
and phonon energy  flux~\cite{galperin_photonics_2017}, $J_P$, out of the system
\begin{equation}
\begin{split}
& A_m(E)=-\frac{1}{\pi}\mbox{Im}\, G^r_{mm}(E)
\\
& I_L=-I_R=\int\frac{dE}{2\pi}\,\mbox{Tr} \left[\Sigma^{<}_{L}(E)\,G^{>}(E)-\Sigma^{B\, >}_{L}(E)\, G^{<}(E)\right]
\\
& J_P=\int\frac{dE}{2\pi}\, E\,\mbox{Tr} \left[\Pi^{<}_{P}(E)\,D^{>}(E)-\Pi^{>}_{P}(E)\, D^{<}(E)\right]
\end{split}
\end{equation}
at steady-state.
Here, $<$, $>$ and $r$ are respectively lesser, greater and retarded projections
of the GFs, self-energies $\Sigma$ and $\Pi$ are defined in Eq.(\ref{SP}), and trace is over molecular orbitals.
in expression for $I_{L\,(R)}$ and over intra-molecular transitions in expression for $J_P$.

Reference system for both models utilizes three auxiliary sites: two mediating coupling of the physical site 
to full and empty Fermi baths and one two-level system mediating coupling between physical site 
and empty Bose bath (see Fig.~\ref{fig1}b and Appendix~\ref{appB}).
As mentioned earlier bigger sizes of auxiliary system require implementation of advanced
methods (e.g., based on MPS) to solve auxiliary QME. Here, we restrict our consideration to 
small sizes which can be evaluated by direct diagonalization of the Liouvillian.
We note that while for such small size representation of physical hybridization function
in the auxiliary system is of limited quality (see Fig.~\ref{figS2}), the aux-DB superperturbation expansion
in the difference of the two hybridization function allows to obtain high quality results even
for small reference system sizes. 

%%%%%%%%%%%%%%%%%%%%%%%%%%%%%%%%%%%%%%%%%%%%%%%%%%%%%%
\begin{figure}[b]
\centering\includegraphics[width=\linewidth]{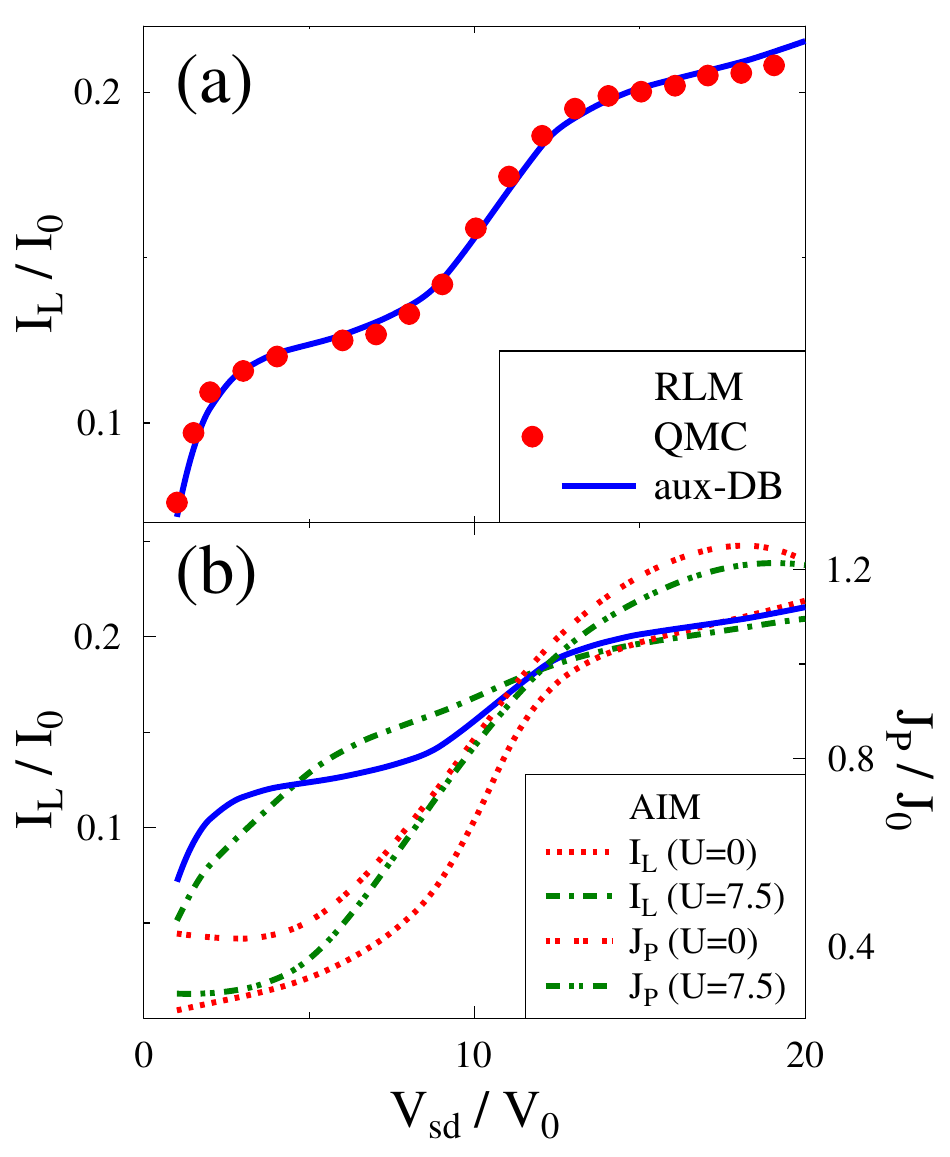}
\caption{\label{fig2}
(Color online) 
Electron $I_L$ and phonon $I_P$ fluxes.
Shown are results for (a) RLM and (b) AIM.
In panel (a) aux-DB results (solid line, blue) are benchmarked vs. numerically exact QMC calculation of 
Ref.~\onlinecite{muhlbacher_real-time_2008}. Panel (b) compares aux-DB results for AIM with $U=0$ 
and $U=7.5$. 
Sold line (blue) presents RLM simulations within aux-DB and is the same in both panels.
}
\end{figure}
%%%%%%%%%%%%%%%%%%%%%%%%%%%%%%%%%%%%%%%%%%%%%%%%%%%%%%

% numerical results
\subsection{Numerical results}
We start from consideration of RLM studied within numerically 
exact QMC approach in Ref.~\onlinecite{muhlbacher_real-time_2008}.
Parameters (in arbitrary energy units $E_0$) are $k_BT=0.2$ and $\varepsilon_0=3.2$.
Following Ref.~\onlinecite{muhlbacher_real-time_2008}
Fermi baths are treated within the wide-band approximation (WBA) with a soft cut-off:
$\Gamma_{L/R}(E)=\Gamma_{L/R}/[1+e^{\nu(E-E_C)}][1+e^{-\nu(E-E_C)}]$ with
 $\nu=5$, $E_C=20$ and $\Gamma_L=\Gamma_R=0.5$;
 Bose bath is characterized by spectral density
 $J(\omega)=\gamma\omega/\big([(\omega/\omega_0)^2-1]^2+[\gamma\omega_0\omega/(2M_0^2)]^2\big)$
 with $\gamma=0.1$, $\omega_0=5$ and $M_0=4$. Bias was applied symmetrically: $\mu_L=-\mu_R=V/2$.
 Results of simulation are presented in terms of units of bias $V_0=E_0/|e|$, flux $I_0=E_0/\hbar$, and energy flux
 $J_0=E_0^2/\hbar$.
 Fig.~\ref{fig2}a compares aux-DB results (solid line) with numerically exact QMC (circles) simulations of 
 Ref.~\onlinecite{muhlbacher_real-time_2008}. 
 
Aux-DB simulations of the AIM with symmetric coupling to Bose bath 
for $U=0$ (dotted line) and $U=7.5$ (dashed line) Coulomb interaction are shown 
 in Fig.~\ref{fig2}b.
 Note that even in the absence of Coulomb interaction results of simulations are significantly different from 
 results of the RLM (compare dotted and solid lines). This is due to effective electron-electron interaction 
 induced by coupling to common Bose bath and the effect can be understood within an effective negative-U model
 ($\tilde \epsilon_0=\epsilon_0-M_0^2/\omega_0$ and $\tilde U=U-2M_0^2/\omega_0$)
 which predicts doubly populated state $E_2=2\tilde\varepsilon_0+\tilde U=-6.4$ to be the ground state of $U=0$ quantum dot  
 with energy gap of $6$ to its singly populated state $E_1=\tilde\varepsilon_0=0$.
 This shows that use of spinless models in studies of inelastic transport
 should be done with caution. For $U=7.5$ (dash-dotted line) no current blockade is observed because electron
 transition from ground state is gapless.
 It is interesting to note that in blockaded region energy (phonon) flux is higher than for resonant tunneling
 (compare double-dotted and dash-double-dotted lines in Fig.~\ref{fig2}b), which indicates predominantly elastic character of resonant transport.

%%%%%%%%%%%%%%%%%%%%%%%%%%%%%%%%%%%%%%%%%%%%%%%%%%%%%%
\begin{figure}[t]
\centering\includegraphics[width=\linewidth]{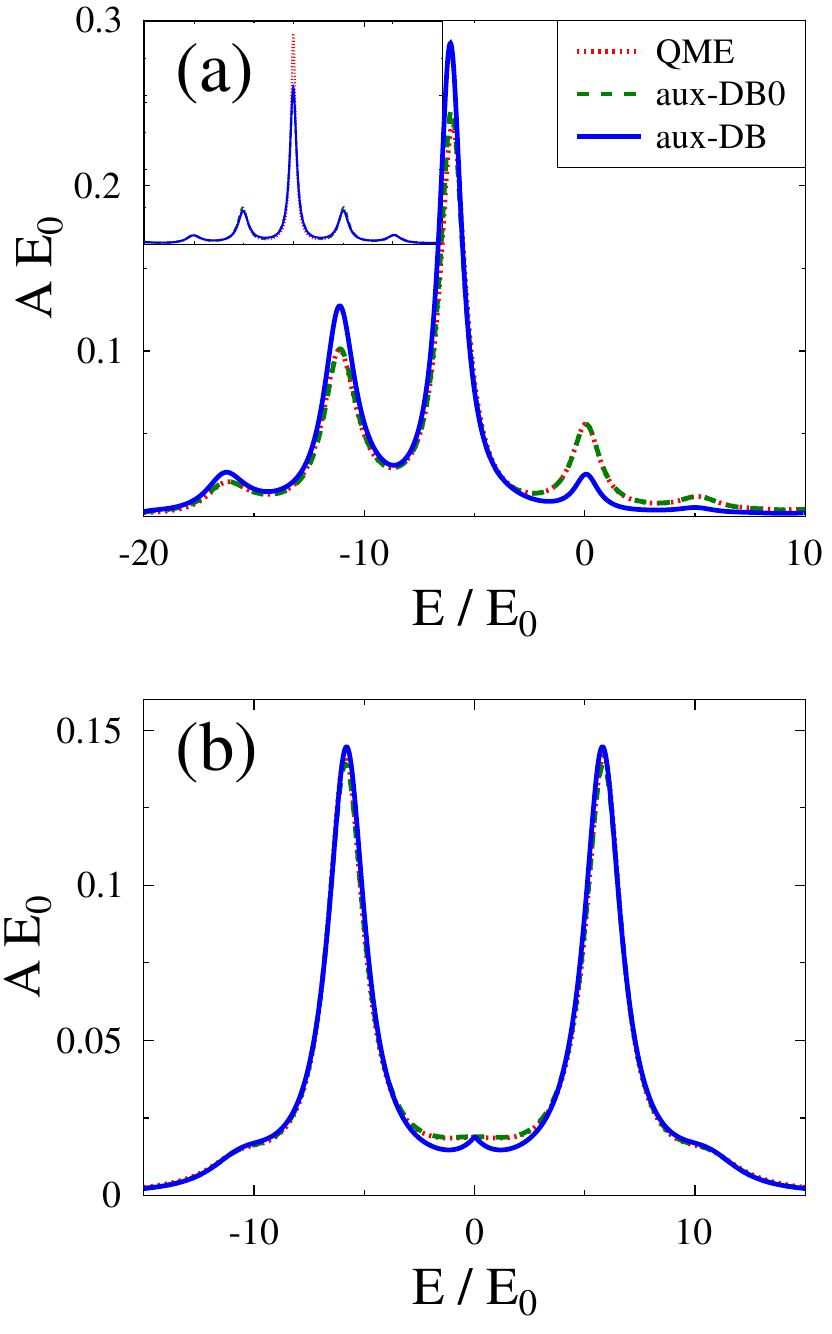}
\caption{\label{fig3}
(Color online) Spectral function $A$ for
AIM with (a) symmetric ($U=0$, $V_{sd}=6$) and (b) anti-symmetric ($U=5$, $V_{sd}=0$) 
couplings to the thermal bath $P$. 
Shown are results of the auxiliary QME (dotted line, red), zero (dashed line, green),
and first order (solid line, blue) aux-DB approaches. 
Inset in (a) shows aux-DB results for RLM.
}
\end{figure}
%%%%%%%%%%%%%%%%%%%%%%%%%%%%%%%%%%%%%%%%%%%%%%%%%%%%%%

Fig.~\ref{fig3} shows spectral functions simulated within the QME (dotted line), zero (dashed line), and
first (solid line) aux-DB approaches for the cases of (a) symmetric and (b) anti-symmetric
couplings to thermal bath. 
Fig.~\ref{fig3}a shows results for AIM with $U=0$, $M_0=4$ and symmetric coupling at $V_{sd}=6$. 
Corresponding RLM results are given in the inset.
While in RLM aux-DB is accurate already in the zero order, AIM $U=0$ results are significantly renormalized
when vertex corrections are taken into account.
Fig.~\ref{fig3}b shows  results for AIM with $k_BT=0$, $U=5$ and $\varepsilon_0=-U/2$, $M_0=0.1$
and anti-symmetric coupling at zero bias. 
One sees, that also in this case vertex corrections are important: they are necessary to reproduce Kondo feature.  

%%%%%%%%%%%%%%%%%%%%%%%%%%%%%%%%%%%%%%%%%%%%%%%%%%%%%%
\begin{figure}[t]
\centering\includegraphics[width=\linewidth]{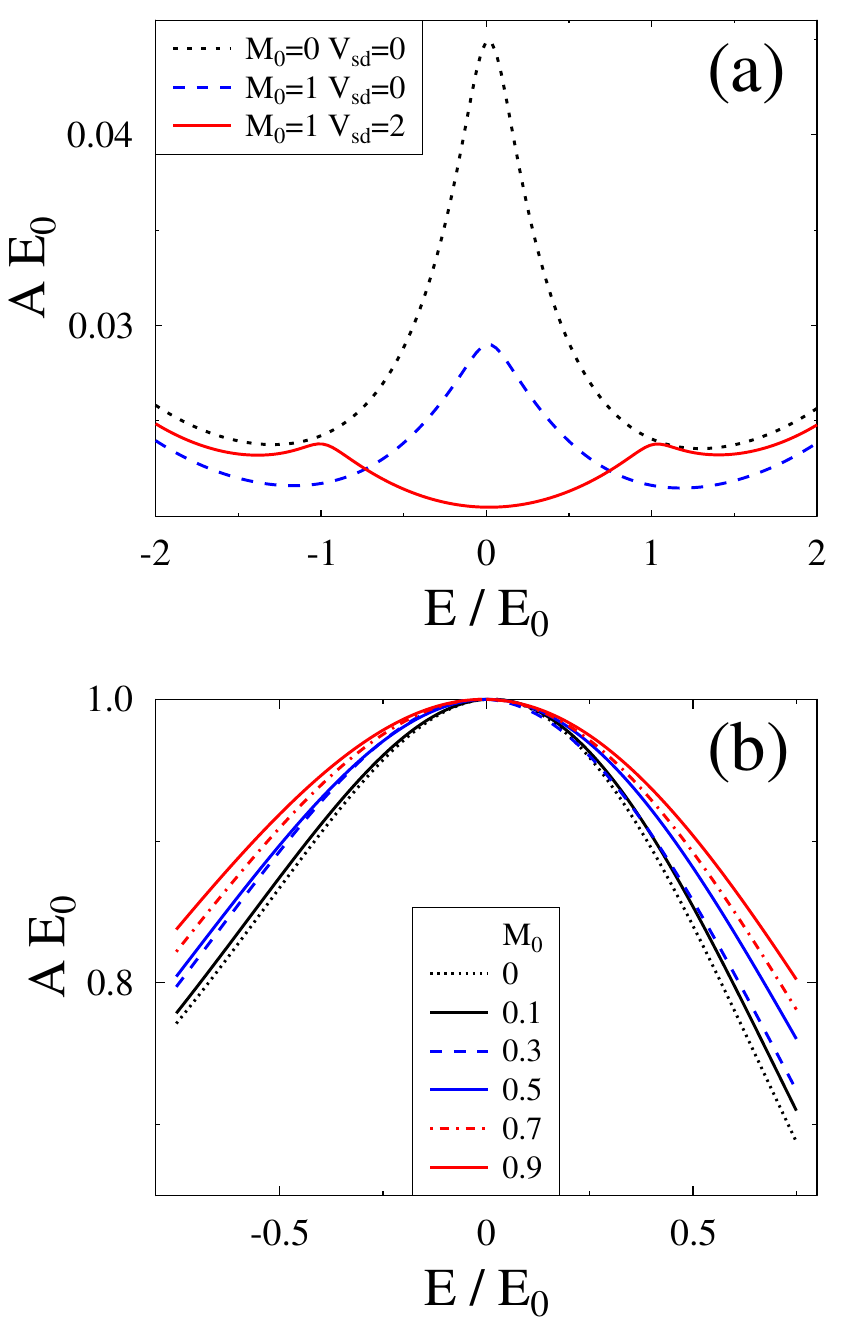}
\caption{\label{fig4}
(Color online)
Spectral function $A$ of AIM with anti-symmetric coupling to thermal bath $P$. 
Shown are results for several molecule-thermal bath coupling strengths 
(a) destruction of the Kondo peak by dephasing induced by coupling to thermal bath
and (b) broadening of the Coulomb peaks (for comparison all Coulomb peaks are shifted to common maximum set at $E=0$).
}
\end{figure}
%%%%%%%%%%%%%%%%%%%%%%%%%%%%%%%%%%%%%%%%%%%%%%%%%%%%%%

Fig.~\ref{fig4}a shows that Kondo is destroyed when increasing coupling strength $M_0$ to the thermal bath
(compare dotted and dashed lines).  The effect is due to the bath induced dephasing.
Nonequilibrium simulation (solid line) shows the Kondo feature splitting.
Finally, in Fig.~\ref{fig4}b we show increase of Coulomb peaks broadening with increase of 
the coupling $M_0$. 
Here parameters are $k_BT=0$, $U=5$ and $\varepsilon_0=-U/2$, so that particle-hole symmetry is fulfilled.
As previously, Fermi baths are considered within the WBA with $\nu=10$ and $E_C=20$.
Bose bath is taken to be Ohmic: $J(\omega)=M_0\, \omega\, e^{-\omega/\omega_C}$ with $\omega_C=20$.
To facilitate comparison peaks are shifted and scaled so that their maxima coincide
and are equal to $1$.

%%%%%%%%%%%%%%%%%%%%%%%%%%%%%%%%%%%%%%%%%%%%%%%%%%%%%%
%%%%%%%%%%%%%%%%%%%%%%%%%%%%%%%%%%%%%%%%%%%%%%%%%%%%%%
% conclusion
\section{Conclusion}\label{conclude}
The nonequilibirum DF approach introduced originally in Ref.~\onlinecite{jung_dual-fermion_2012}
and its optimization for steady-state simulations - the aux-DF approach~\cite{chen_auxiliary_2019} -
are promising methods for modeling strongly correlated open systems.
Contrary to usual diagrammatic expansions the methods can treat systems with no small parameter available.
This is the situation often encountered in single-molecule optoelectronic devices, which are at the forefront
of experimental and theoretical research due to interesting fundamental problems and applicational perspectives
in energy nano-materials, spintronics, and quantum computation.
However, application of the aux-DF to simulations of optoelectronic devices is hindered by its inability 
to account for energy exchange between molecule and plasmonic field. The latter is crucial in modeling of the devices. 

Here we proposed a new nonequilibrium method, {\em the aux-DB approach},
which accounts for both electron and energy fluxes between system and baths.
The nonequilibrium aux-DB is a super-perturbation theory inspired by
equilibrium DB method~\cite{rubtsov_dual_2012} proposed as generalization of the extended DMFT.
Employing auxiliary QME and choosing infinite reference system
makes the approach advantageous in treating the steady-states.

We utilized generic junction models of a molecule coupled to two Fermi leads and Bose phonon bath.
The aux-DB was benchmarked vs. numerically exact QMC results of Ref.~\onlinecite{muhlbacher_real-time_2008}.
We showed that the new scheme is both accurate and relatively numerically inexpensive.
Further development of the method and its application to realistic systems is a goal for future research.

%%%%%%%%%%%%%%%%%%%%%%%%%%%%%%%%%%%%%%%%%%%%%%%%%%%%%%%%%%%%%%%%%%%%%%%%%%%%%%%

\begin{acknowledgments}
M.G. acknowledges support by the National Science Foundation  (grant CHE-1565939).
The work of M.I.K. is supported by European Research Council via Synergy
Grant 854843 - FASTCORR
\end{acknowledgments}

%%%%%%%%%%%%%%%%%%%%%%%%%%%%%%%%%%%%%%%%%%%%%%%%%%%%%%%%%%%%%%%%%%%%%%%%%%%%%%%

\appendix

\section{Derivation of dual boson EOMs}\label{appA}
Here we present derivation of the expressions for the zero order GFs, $G_0^{DF}$ and $D_0^{DB}$, and
self-energies, $\Sigma^{DF}$ and $\Pi^{DB}$, for the dual boson technique, Eq.~(6) of the main text.

We consider a physical system which consists from the molecule ($d$)
coupled to Fermi ($c$) and Bose ($a$) baths. Its partition function on the Keldysh contour is~\cite{kamenev_field_2011}
\begin{equation}
\label{Z}
 Z=\int_c D[\bar d,d,\bar c,c,\bar a,a]\, e^{iS[\bar d,d,\bar c,c,\bar a,a]}
\end{equation}  
where
\begin{equation}
\begin{split}
 & S[\bar d,d,\bar c,c,\bar a,a] = \bar d_1\, \left[G_0^{-1}\right]_{12} d_2  +S^{int}[\bar d,d]
 \\ &
  + \bar c_1 \left[g_B^{-1}\right]_{12} c_2
  + \bar a_1 \left[d_B^{-1}\right]_{12} a_2
  \\
 & + \bar d_1 V_{12} c_2 + \bar c_2 V_{21} d_1 
    + \bar b_1 V_{12} a_2 + \bar a_2 V_{21} b_1
 \end{split}
\end{equation}
is the action of an interacting system (molecule) coupled to non-interacting contacts (Fermi bath)
and plasmon (Bose bath).
Here, $G_0^{-1}$ is defined in Eq.~(2) of the main text and $g_B^{-1}$ and $d_B^{-1}$ are the inverse GFs 
for free electrons in the contacts and free photons in the Bose bath
\begin{equation}
 \begin{split}
 \left[g_B^{-1}\right]_{12} &= \delta(\tau_1,\tau_2)\left[i\partial_{\tau_1}-\varepsilon_k\right]
 \\
 \left[d_B^{-1}\right]_{12} &= \delta(\tau_1,\tau_2)\left[i\partial_{\tau_1}-\omega_\alpha\right]
 \end{split}
\end{equation}
After integrating out baths degrees of freedom~\cite{negele_quantum_1988} one gets 
effective action presented in Eq.~(1) of the main text.

Next we introduce {\em an exactly solvable reference system}, 
which is identical to the original one in all intra-system interactions
but differs from it by its hybridization function. Effective action of the original system will be related 
to that of the reference system via Eq.~(4) of the main text. 
Because direct application of perturbation theory to Eq.~(4) is not possible, we apply two Hubbard-Stratonovich 
transformations to introduce new particles, {\em dual fermion} ($f$) and {\em dual boson} ($\eta$),
which disentangle last two terms in Eq.~(4). Following Ref.~\onlinecite{van_loon_beyond_2014} we get
\begin{equation}
\begin{split}
& e^{\bar d_1 N_{12} d_2} = 
\\ &\ \
Z_f\,\int_c D[\bar f,f] \,
e^{-\bar f_1 \alpha^f_{12} \left[N^{-1}\right]_{23}\alpha^f_{34} f_4 + \bar f_1 \alpha^f_{12} d_2 + \bar d_1 \alpha^f_{12} f_2}
\\
& e^{\bar b_1 M_{12} b_2} = 
\\ & \ \
Z_b\,\int_c D[\bar \eta,\eta] \,
e^{-\bar \eta_1 \alpha^b_{12} \left[M^{-1}\right]_{23}\alpha^b_{34} \eta_4 + \bar \eta_1 \alpha^b_{12} b_2 + \bar b_1 \alpha^b_{12} \eta_2}
\end{split}
\end{equation}
with 
\begin{equation}
\begin{split}
 &\alpha^f = i\, g^{-1} \quad N = i\, \delta\Sigma^B \quad Z_f=\big(\det{\left[\alpha^f\, N^{-1}\,\alpha^f\right]}\big)^{-1}
 \\
 &\alpha^b = i\,\chi^{-1} \quad M=i\,\delta\Pi^B \quad Z_b=\det{\left[\alpha^b\, M^{-1}\,\alpha^b\right]}
\end{split}
\end{equation}
Applying the transformation to the partition function (\ref{Z}) with the action given by Eq.~(4) of the main text yields
\begin{equation}
\label{ZLeft}
Z = Z_f\, Z_b\, \int_c D[\bar d,d,\bar f,f,\bar\eta,\eta] e^{iS[\bar d,d,\bar f,f,\bar\eta,\eta]}
\end{equation}
where
\begin{equation}
\label{SLeft}
\begin{split}
 & S[\bar d,d,\bar f,f,\bar\eta,\eta] = \tilde S[d^*,d] 
 \\ &
  -\bar f_1\, g^{-1}_{12}\,\left[\delta\Sigma^B\right]^{-1}_{23}\, g^{-1}_{34}\, f_4 
  +\bar f_1\, g^{-1}_{12}\, d_2 + \bar d_1\, g_{12}^{-1} f_2 
  \\ &
  -\bar\eta_1\chi^{-1}_{12}\,\left[\delta\Pi^B\right]^{-1}_{23}\,\chi^{-1}_{34}\,\eta_4
  +\bar\eta_1\,\chi^{-1}_{12}\, b_2 + \bar b_1\,\chi^{-1}_{12}\,\eta_2
 \end{split}
\end{equation}
Thus, auxiliary quasi-particles - dual fermion ($f$) and dual boson ($\eta$) - were introduced.

Integrating out of the real quasiparticle, $\bar d$ and $d$, in (\ref{ZLeft}) leads to
\begin{equation}
\label{ZRight}
Z = Z_f\, Z_b\,\tilde Z \int_c D[\bar f,f,\bar\eta,\eta] e^{iS[\bar f,f\bar\eta,\eta] }
\end{equation}
with
\begin{equation}
\begin{split}
S[\bar f,f\bar\eta,\eta] &= \bar f_1\,\left[G_0^{DF}\right]^{-1}_{12}\, f_2 +
\bar\eta_1\,\left[D_0^{DB}\right]^{-1}_{12}\,\eta_2
\\ & + V[\bar f,f,\bar\eta,\eta]
\end{split}
\end{equation}
$\left[G_0^{DF}\right]^{-1}_{12}$ and $\left[D_0^{DB}\right]^{-1}_{12}$ are defined in Eq.~(6) of the main text,
$\tilde Z$ is the partition function of the reference system, and $V[\bar f,f,\bar\eta,\eta]$ is
unknown interaction between dual particles.

%%%%%%%%%%%%%%%%%%%%%%%%%%%%%%%%%%%%%%%%%%%%
\begin{figure*}[htbp]
\centering\includegraphics[width=\linewidth]{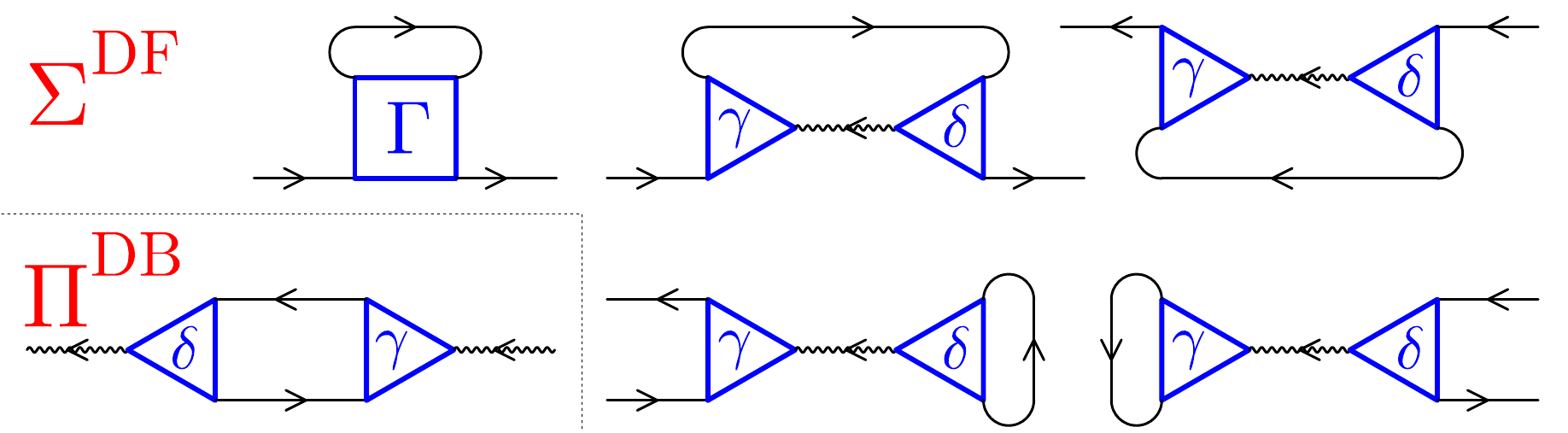}
\caption{\label{figS1}
Contributions to diagrams for dual fermion, $\Sigma^{DF}$, and dual boson, $\Pi^{DB}$, self-energies, Eq.~(6).
Directed solid and wavy lines (black) indicate dual fermion and dual boson GFs, $G_0^{DF}$ and
$D_0^{DB}$, respectively.
Triangle and square (blue) indicate vertices $\gamma$ and $\Gamma$ of the reference system.
}
\end{figure*}
%%%%%%%%%%%%%%%%%%%%%%%%%%%%%%%%%%%%%%%%%%%%

To get the interaction $V[\bar f,f,\bar\eta,\eta]$ we expand (\ref{ZLeft}) in $f-d$ and $\eta-b$ interactions 
and integrate out real quasiparticles, $\bar d$ and $d$.
Taking $g$ and $\chi$ to be single electron and single molecular excitaton GFs of the reference system 
\begin{equation}
\label{gchi}
\begin{split}
 g_{12} &= \frac{-i}{\tilde Z}\int_c D[\bar d,d]\, d_1 \bar d_2 \, e^{i \tilde S[\bar d,d]} 
  \equiv -i\langle T_c\, \hat d_1\, \hat d_2^\dagger\rangle_{ref}
 \\
 \chi_{12} &= \frac{-i}{\tilde Z}\int_c D[\bar d,d]\, \delta b_1\, \delta\bar b_2 \, e^{i \tilde S[\bar d,d]}
 \equiv -i \langle T_c\, \hat b_1\,\hat b_2^\dagger\rangle_{ref}
\end{split}
\end{equation}
and comparing the resulting expression to expansion of (\ref{ZRight}) yields expression for $V[\bar f,f,\bar\eta,\eta]$.
In particular, for expansion up to fourth order in $\bar f$, $f$ and second order in $\bar\eta$, $\eta$
\begin{equation}
\label{V}
\begin{split}
 & V[\bar f,f,\bar\eta,\eta] = \bar\eta_1\,\chi^{-1}_{12}\,\langle b_2\rangle_{ref} + \langle\bar b_1\rangle_{ref}\,\chi^{-1}_{12}\,\eta_2
 \\ &
 -\frac{i}{4} \bar f_1\,\bar f_3 \Gamma_{13;24} f_2 f_4
 \\ &
 -\bar\eta_1\,\gamma_{123}\,\bar f_2\, f_3 - \bar f_3\, f_2\,\delta_{321}\, \eta_1
 \end{split}
\end{equation}
Here $\gamma_{123}$, $\delta_{321}$ and $\Gamma_{13;24}$ are vertices of the reference system 
\begin{equation}
\label{vertices}
\begin{split}
  \Gamma_{13;24} &=
  g^{-1}_{11'}\, g^{-1}_{33'}\,
  \big[-\langle T_c\, \hat d_{1'}\,\hat d_{2'}^\dagger\,\hat d_{3'}\,\hat d_{4'}^\dagger\rangle_{ref}
  \\ &
  -g_{1'2'}\,g_{3'4'}+g_{1'4'}\, g_{3'2'}\big]\, g^{-1}_{2'2}\, g^{-1}_{4'4}
  \\
  \gamma_{123} &= \chi^{-1}_{11'}\,g^{-1}_{22'}\,\langle T_c\,\delta \hat b_{1'}\,\hat d_{2'}\,\hat d_{3'}^\dagger \rangle_{ref}\, g^{-1}_{3'3}
  \\
  \delta_{321} &= g^{-1}_{33'}\,\langle T_c\,\hat d_{3'}\,\hat d_{2'}^\dagger\,\delta\hat b_{1'}^\dagger\rangle_{ref}\,\chi^{-1}_{1'1}\, g^{-1}_{2'2} 
\end{split}
\end{equation}
Here $T_c$ is contour ordering operator, subscript $ref$ indicates Markov Lindblad-type evolution of 
the reference system and $\delta\hat b\equiv \hat b - \langle\hat b\rangle_{ref}$.
We note in passing that projections of the vertices $\gamma_{123}$ and $\delta_{321}$ are related via
\begin{equation}
\left[\gamma_{123}^{s_1s_2s_3}\right]^{*} = -\delta_{321}^{\bar s_3\bar s_2\bar s_1}
\end{equation}
where $s_{1,2,3}\in\{-,+\}$ indicate branches of the Keldysh contour and $\bar s$ is the branch opposite to s.

Finally, using (\ref{ZRight}) with interaction given by (\ref{V}) in expansion of GFs for the dual particles 
\begin{equation}
\begin{split}
G_{12} &\equiv -i\langle T_c\, f_1\,\bar f_2\rangle
\\
D_{12} &\equiv -i\langle T_c\, b_1\,\bar b_2\rangle
\end{split}
\end{equation}
up to second order and employing the Wick's theorem yields the dual particles self-energies given in Eq.~(6) of the main text.
Corresponding diagrams are shown in Fig.~\ref{figS1}.

%%%%%%%%%%%%%%%%%%%%%%%%%%%%%%%%%%%%%%%%%%%%%%%%%%%%%%%%%%%%%%%%%%%

%%%%%%%%%%%%%%%%%%%%%%%%%%%%%%%%%%%%%%%%%%%%
\begin{figure*}[htbp]
\centering\includegraphics[width=\linewidth]{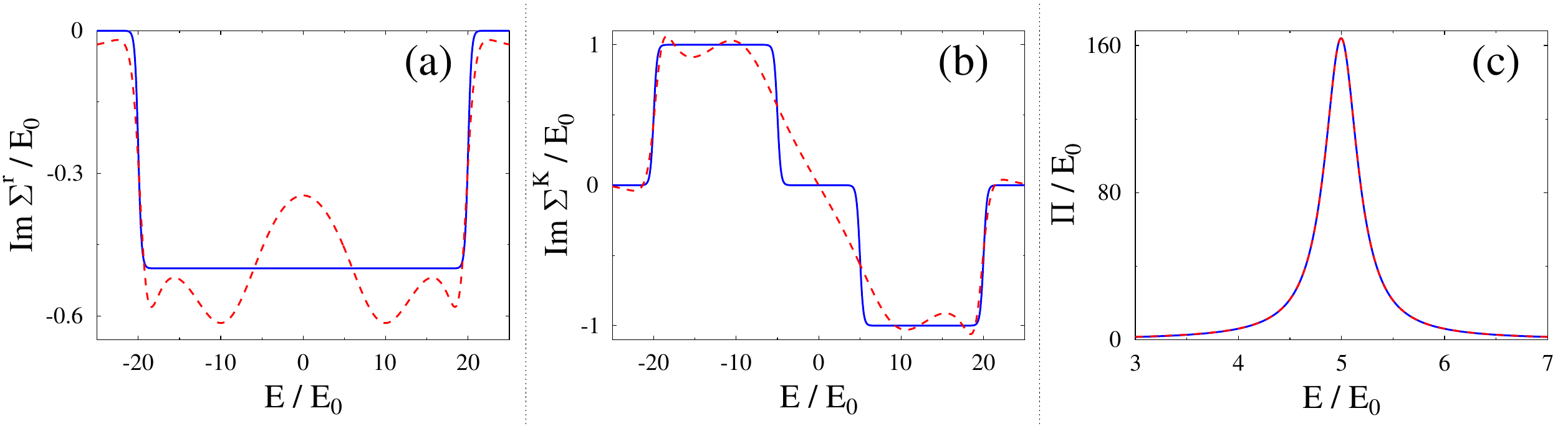}
\caption{\label{figS2}
Hybridization functions of the physical (solid line, blue) and auxiliary (dashed line, red) systems.
Shown are (a) retarded and (b) Keldysh projections of the self-energy due to coupling to contacts and
(c) hybridization function due to coupling to thermal bath. 
Fitting is done for parameters adopted in the first numerical example presented in the main text.
}
\end{figure*}
%%%%%%%%%%%%%%%%%%%%%%%%%%%%%%%%%%%%%%%%%%%%

\section{Fitting hybridization functions with auxiliary modes}\label{appB}
Recently, exact proof of possibility to map unitary evolution of a physical system onto Markov Lindblad-type evolution of an 
auxiliary system was established for systems interacting with Fermi~\cite{arrigoni_nonequilibrium_2013,dorda_auxiliary_2015,dorda_optimized_2017,chen_markovian_2019} and Bose~\cite{tamascelli_nonperturbative_2018,mascherpa_optimized_2019} baths.  
At the heart of the mapping is fitting of hybridization functions of the physical system with set of auxiliary modes in the
auxiliary system. Here, we give details of the fitting procedure.

Explicit form for the Markov Lindblad-type QME (9) is
\begin{equation}
\label{QME}
\frac{d\rho^{SA}(t)}{dt}=-i\mathcal{L}\rho^{SA}(t) \equiv -i[\hat{H}_{SA},\rho^{SA}(t)]+\mathcal{D}\rho^{SA}(t)
\end{equation}
with the Liouvillian taken as
\begin{equation}
\begin{split}
\label{Liouv}
& \hat{H}_{SA} = 
\hat{H}_S +\sum_{n_1,n_2}\epsilon_{m_1m_2}\hat c^\dagger_{n_1} \hat c_{n_2}
\\ &
+\sum_{m,n}\big(t_{mn}\hat d_m^\dagger \hat c_n+t_{mn}^{*}\hat c_n^\dagger\hat d_m\big) 
\\ &
+\sum_{\beta_1,\beta_2}\omega_{\beta_1\beta_2}\hat e_{\beta_1}^\dagger \hat e_{\beta_2}
\\ &
+\sum_{m_1,m_2,\beta}r_{m_1m_2}^{\beta} (\hat b^\dagger_{m_1m_2} +\hat b_{m_1m_2}^\dagger)
(\hat e_\beta+\hat e_\beta^\dagger)
\\
\mathcal{D}\rho &=
\sum_{n_1,n_2}\bigg(\Gamma^{(R)}_{n_1n_2}\big(\hat c_{n_2}\,\hat \rho\,\hat c^\dagger_{n_1}
-\frac{1}{2}\{\hat \rho,\hat c^\dagger_{n_1}\hat c_{n_2}\}\big)
\\ &\qquad\quad
+\Gamma^{(L)}_{n_1n_2}\big(\hat c_{n_1}^\dagger\,\hat\rho\,\hat c_{n_2}
-\frac{1}{2}\{\hat\rho,\hat c_{n_2} \hat c^\dagger_{n_1}\}\big)\bigg)
\\ &
+\sum_{\beta_1,\beta_2}\gamma^{(P)}_{\beta_1\beta_2}\big(\hat e_{\beta_2}\,\hat \rho\,\hat e^\dagger_{\beta_1}
-\frac{1}{2}\{\hat e^\dagger_{\beta_1}\hat e_{\beta_2},\hat\rho\}\big)
\end{split}
\end{equation}
Here $\hat c_n^\dagger$ ($\hat c_n$) and $\hat e^\dagger_\beta$ ($\hat e_\beta$)
create (annihilate) excitation in auxiliary Fermi mode $n$ and Bose mode $\beta$, respectively.

Following Refs.~\onlinecite{dorda_auxiliary_2015,chen_markovian_2019} we construct 
retarded, $\tilde\Sigma^r$, and Keldysh, $\tilde\Sigma^K$, projections of the Fermi hybridization function
in the auxiliary system as
\begin{equation}
\label{tildeSigmaE}
\begin{split}
\tilde{\Sigma}_{m_1m_2}^{r}(E)&=\sum_{n_1,n_2}t_{m_1n_1}\,\tilde G^{r}_{n_1n_2}(E)\, t_{m_2n_2}^{*}
\\
\tilde{\Sigma}_{m_1m_2}^{K}(E)&=\sum_{n_1,n_2}t_{m_1n_1}\,\tilde G^{K}_{n_1n_2}(E)\, t_{m_2n_2}^{*}
\end{split}
\end{equation}
where
\begin{equation}
\begin{split}
\tilde{\mathbf{G}}^r(E) &= \bigg(E\,\mathbf{I}-\mathbf{\epsilon}+\frac{i}{2}(\mathbf{\Gamma}^{(R)}+\mathbf{\Gamma}^{(L)})\bigg)^{-1}
\\
\tilde{\mathbf{G}}^K(E) &=  i\, \tilde{\mathbf{G}}^r(E)\,\big(\mathbf{\Gamma}^{(L)}-\mathbf{\Gamma}^{(R)}\big)\,\tilde{\mathbf{G}}^a(E)
\end{split}
\end{equation}
are retarded, $\tilde{\mathbf{G}}^r(E)$, and Keldysh, $\tilde{\mathbf{G}}^K(E)$ projections of the Fermi auxiliary modes Green's functions,  and where $\tilde{\mathbf{G}}^a(E)\equiv[\tilde{\mathbf{G}}^r(E)]^\dagger$ is its advanced projection.
Hybridization functions (\ref{tildeSigmaE}) should fit corresponding hybridization functions
\begin{equation}
\label{SigmaE}
\begin{split}
\Sigma^r_{m_1m_2}(E) &= \sum_{k\in \{L,R\}} V_{m_1k}\, g^r_k(E)\, V_{km_2}
\\
\Sigma^K_{m_1m_2}(E) &= \sum_{k\in \{L,R\}} V_{m_1k}\, g^K_k(E)\, V_{km_2}
\end{split}
\end{equation}
of the physical system. Here
\begin{equation}
\begin{split}
 g_k^r(E) &\equiv \big(E-\varepsilon_k+i\delta\big)^{-1}
 \\
 g_k^K(E) &\equiv 2\pi i\, (2 n_k -1 )\delta(E-\varepsilon_k)
\end{split}
\end{equation}
are the retarded and Keldysh projections of the free electron in state $k$ in contacts,
$n_k$ is the Fermi-Dirac thermal distribution and $\delta=0^{+}$

We construct Bose hybridization function in the auxiliary system following Refs.~\onlinecite{tamascelli_nonperturbative_2018,mascherpa_optimized_2019}.
For the physical system-bosonic bath coupling taken in the form
\begin{equation}
\sum_{m_1m_2}\sum_\alpha\, V_{m_1m_2}^\alpha (\hat b_{m_1m_2}+\hat b_{m_1m_2}^\dagger)
(\hat{a}_\alpha+\hat{a}_\alpha^\dagger)
\end{equation}
the effect of the bosonic environment can be fully encoded by correlation function
\begin{equation}
\begin{split}
\label{Pi}
&{\Pi}_{m_1m_2,m_3m_4}(t-t') =
\\ &\quad
 \sum_\alpha V_{m_1m_2}^\alpha
\braket{(\hat a_\alpha+\hat{a}_\alpha^\dagger)(t)\,(\hat{a}_\alpha+\hat{a}_\alpha^\dagger)(t')}
V_{m_3m_4}^\alpha
\end{split}
\end{equation}
Similarly, coupling to auxiliary Bose modes in (\ref{Liouv}) is fully described by correlation function
\begin{equation}
\begin{split}
\label{tildePi}
&\tilde{\Pi}_{m_1m_2,m_3m_4}(t-t') =
\\ &\quad
 \sum_{\beta_1,\beta_2} r_{m_1m_2}^{\beta_1}
\braket{(\hat e_{\beta_1}+\hat{e}_{\beta_1}^\dagger)(t)\,(\hat{e}_{\beta_2}+\hat{e}_{\beta_2}^\dagger)(t')}
r_{m_3m_4}^{\beta_2}
\\ &\quad
\equiv i\sum_{\beta_1,\beta_2}r_{m_1m_2}^{\beta_1}\big(\tilde D_{\beta_1\beta_2}^{>}(t-t')+
\tilde D_{\beta_2\beta_1}^{<}(t'-t)\big)
\end{split}
\end{equation}
Here $\tilde D^{>}$ and $\tilde D^{<}$ are the greater and lesser projections of Bose auxiliary mode
Green's function
\begin{equation}
\label{tildeD}
 \tilde D_{\beta_1\beta_2}(\tau_1,\tau_2) = -i\langle T_c\,\hat e_{\beta_1}(\tau_1)\,\hat e_{\beta_2}^\dagger(\tau_2)\rangle
\end{equation}
Fourier transform of the correlation function (\ref{tildePi}) is
\begin{equation}
\label{tildePiE}
\tilde{\Pi}_{m_1m_2,m_3m_4}(E) =
i\sum_{\beta_1,\beta_2}r_{m_1m_2}^{\beta_1}\big(\tilde D_{\beta_1\beta_2}^{>}(E)+
\tilde D_{\beta_2\beta_1}^{<}(-E)\big)
\end{equation}
According to Ref.~\onlinecite{tamascelli_nonperturbative_2018} in auxiliary system one considers 
Bose bath at zero temperature with eigenmodes spanning energy range from -$\infty$ to $+\infty$. 
Thus, greater and lesser projections of the Green's function (\ref{tildeD}) satisfy
\begin{equation}
\begin{split}
\tilde{\mathbf{D}}^{>}(E) &= -i\,\tilde{\mathbf{D}}^r(E)\,\mathbf{\gamma}^{(P)}\,\tilde{\mathbf{D}}^a(E)
\\
\tilde{\mathbf{D}}^{<}(E) &=0
\end{split}
\end{equation}
where
\begin{equation}
\begin{split}
\tilde{\mathbf{D}}^r(E) &= \big(E\, \bm{I}-\mathbf{\omega}+\frac{i}{2}\mathbf{\gamma}^{(P)}\big)^{-1}
\\
\tilde{\mathbf{D}}^a(E) &= \big[\tilde{\mathbf{D}}^r(E)\big]^\dagger
\end{split}
\end{equation}
are the retarded projection and advanced projections.

For the correlation function (\ref{Pi}) representing physical system and for the case of thermal Bose bath
with inverse temperature $\beta$
\begin{align}
\label{PiE}
&\Pi_{m_1m_2,m_3m_4}(E) = \bigg(1+\coth\frac{\beta E}{2}\bigg)
\\ & \times
\bigg(J_{m_1m_2,m_3m_4}(E)\,\theta(E)-J_{m_3m_4,m_1m_2}(-E)\,\theta(-E)\bigg)
\nonumber
\end{align}
where 
\begin{equation}
J_{m_1m_2,m_3m_4}(E)\equiv \pi\sum_\alpha V_{m_1m_2}^{\alpha}\, V_{m_3m_4}^{\alpha}\,\delta(E-\omega_\alpha)
\end{equation}
Following Ref.~\onlinecite{mascherpa_optimized_2019} we stress that although the auxiliary Bose bath is taken at zero 
temperature this does not restrict the temperature of Bose bath in the physical system:
the information about finite temperature will be provided by parameters of the auxiliary Bose modes. 

Finally note that parameters $\epsilon_{m_1m_2}$, $t_{mn}$, $\omega_{\beta_1\beta_2}$,
$r^\beta_{m_1m_2}$, $\Gamma^{(L)}_{n_1n_2}$, $\Gamma^{(R)}_{n_1n_2}$ and $\gamma^{(P)}_{\beta_1\beta_2}$
of the Lindblad equation (\ref{QME})-(\ref{Liouv}) 
are used to fit hybridization functions (\ref{SigmaE}) and (\ref{PiE}) of the physical system
with corresponding hybridization functions (\ref{tildeSigmaE}) and (\ref{tildePiE}) of the auxiliary model 
employing a cost function to quantify deviation~\cite{chen_markovian_2019}.
Figure~\ref{figS2} shows hybridization functions for the physical model (solid lines) and their fitting
with auxiliary modes (dashed lines) as utilized in simulations of the RLM and AIM with symmetric coupling to thermal bath 
presented in the main text.  We used four Fermi and one Bose auxiliary modes to fit the corresponding hybridization functions.

%%%%%%%%%%%%%%%%%%%%%%%%%%%%%%%%%%%%%%%%%%%%%%%%%%%%%%%%%%%%%%%%%%%

\section{Green's Functions and vertices of the reference system}\label{appC}
To evaluate dual-particles self-energies, Eq.~(6) of the main text, one has to calculate 
GFs $g$ and $\chi$, Eq.~(\ref{gchi}), and verticies $\gamma$, $\delta$ and $\Gamma$, Eq.~(\ref{vertices}), of the reference system. 
These quantities are given by two- ($g$ and $\chi$), three- ($\gamma$, $\delta$) and four-time ($\Gamma$) correlation functions 
defined on the Keldysh contour. 

To provide these we utilize the quantum regression relation~

Because Markov Lindblad-type QME is sued to solve the reference system, we can employ
the quantum regression relation~\cite{breuer_theory_2003}
\begin{align}
\label{qrr}
& \big\langle T_c\, \hat A(\tau_1)\,\hat B(\tau_2)\ldots \hat Z(\tau_n)\big\rangle =
\\ &\qquad
 \mbox{Tr}\big[\mathcal{O}_n\,\mathcal{U}(t_n,t_{n-1})\ldots
 \mathcal{O}_2\,\mathcal{U}(t_2,t_1)\,\mathcal{O}_1\,\mathcal{U}(t_1,0)\,\rho^{SA}(0)\big]
 \nonumber
\end{align}
to evaluate correlation functions.
Here $\rho^{SA}(0)$ is the steady-state density matrix of the extended system,
$\mathcal{U}(t_i,t_{i-1})$ is the Liouville space evolution operator and
times $t_i$ are ordered so that $t_n>t_{n-1}>\ldots>t_2>t_1>0$.
$\mathcal{O}_i$ is the Liouville space super-operator corresponding to 
one of operators  $\hat A\ldots \hat Z$ whose time is $i$-th in the ordering.
It acts from the left (right) for the operator on the forward (backward) branch of the contour.
The steady-state density matrix is found as a right eigenvector $\lvert R_0\gg$ corresponding to the Liouvillian eigenvalue 
$\lambda_0=0$.
Using spectral decomposition of the Liouvillian, the evolution operator can be presented in its eigenbasis as
\begin{equation}
\label{U}
 \mathcal{U}(t_i,t_{i-1}) = \sum_\gamma \lvert R_\gamma\gg\, e^{-i\lambda_\gamma(t_i-t_{i-1})}\,
 \ll L_\gamma\rvert.
\end{equation}
For evaluation of single- and two-particle GFs,
besides the $\mathcal{L}$ of Eq.~(9) of the main text we will also need
Liouvillians $\mathcal{L}^{(\pm 1)}$ and $\mathcal{L}^{(\pm 2)}$.
These are evolution operator generators for Liouville space vectors $\lvert S_1S_2\gg$ with different 
number $N_S$ of electrons in states $\lvert S_1\rangle$ and $\lvert S_2\rangle$: $N_{S_1}-N_{S_2}=\pm1,\pm2$.

Using (\ref{U}) in (\ref{qrr}) yields expressions for the single-particle ($g$ and $\chi$) and two-particle GFs of 
the reference system (see Appendix~\ref{appC} for details).
To do so we have to consider several projections (contour orderings) and time orderings.
In particular, evaluation of two-time correlation functions requires consideration of $2^1=2$ projections with
$2!=2$ time orderings for each projection.  Three-time correlation functions will require consideration of
$2^2=4$ projections with $3!=6$ time orderings.  Evaluation of four-time correlation function requires
consideration of $2^3=8$ projections with $4!=24$ time orderings. 
Evaluating projections one has to take care of sign of Fermi operators permutations.
\begin{equation}
\begin{split}
 &\langle T_c\, \hat O_1(\tau_1)\,\hat O_2(\tau_2)\ldots\hat O_N(\tau_N)\rangle_{ref}
 =
 \\ &\quad (-1)^P
 \langle\langle I\vert \mathcal{O}_{\theta_1}\,\mathcal{U}(t_{\theta_1},t_{\theta_2})\, \mathcal{O}_{\theta_2}\,
 \mathcal{U}(t_{\theta_2},t_{\theta_3})\ldots 
 \\ &\qquad\qquad\qquad\qquad\ldots
 \mathcal{O}_{\theta_N}\mathcal{U}(t_{\theta_N},0)
 \vert \rho^{SA}(0)\rangle\rangle
 \end{split}
\end{equation}
 Here $P$ is number of Fermi interchanges in the permutation of operators $\hat O_i$ by $T_c$, 
 $\langle\langle I\rvert$ is the Liouville space bra representation of the Hilbert space identity operator,
 $\theta_i$ are indices of operators $\hat O_i$ rearranged is such a way that 
 $t_{\theta_1}>t_{\theta_2}>\ldots >t_{\theta_N}$
 ($t_{\theta_i}$ is real time corresponding to contour variable $\tau_{\theta_i}$),
 $\mathcal{U}$ is the Liouville space evolution superoperator defined in Eq.~(11), and
 $\mathcal{O}_{\theta_i}$ are the Liouville space superoperators corresponding to the Hilbert space operators $\hat O_i$
\begin{equation}
 \mathcal{O}_i \vert \rho\rangle\rangle = 
 \begin{cases}
  \mathcal{O}_i^{-} \vert \rho\rangle\rangle \equiv \hat O_i\,\hat\rho & \mbox{forward branch} 
  \\
  \mathcal{O}_i^{+} \vert \rho\rangle\rangle \equiv \hat\rho\,\hat O_i & \mbox{backward branch} 
 \end{cases}
\end{equation}
Further details on evaluation of multi-time correlation functions can be found in Ref.~\onlinecite{chen_auxiliary_2019}.

Once single- and two-particle GFs of the reference system are known, 
the vertices required in  Eq.~(6) of the main text can be calculated from their definitions, Eq.~(\ref{vertices}).

%%%%%%%%%%%%%%%%%%%%%%%%%%%%%%%%%%%%%%%%%%%%%%%%%%%%%%%%%%%%%%%%%%%%%%%%%%%%%%%
%\bibliography{aux_db}
%%%%%%%%%%%%%%%%%%%%%%%%%%%%%%%%%%%%%%%%%%%%%%%%%%%%%%%%%%%%%%%%%%%%%%%%%%%%%%%

\end{document}